\documentclass[preprint,floatfix,showpacs,aps,nofootinbib]{revtex4}
\usepackage{graphicx}
\usepackage{dcolumn}
\usepackage{bm}
\usepackage{amsmath,amssymb}
\newcommand {\be}      {\begin{equation}}   
\newcommand {\ee}      {\end{equation}}
\newcommand {\bee}     {\begin{equation*}}   
\newcommand {\eee}     {\end{equation*}}
\newcommand {\bea}     {\begin{eqnarray}}   
\newcommand {\eea}     {\end{eqnarray}}
\newcommand {\beaa}    {\begin{eqnarray*}}   
\newcommand {\eeaa}    {\end{eqnarray*}}
\newcommand {\bse}     {\begin{subequations}}
\newcommand {\ese}     {\end{subequations}}

\newcommand {\lam}     {\lambda}
\newcommand {\sig}     {\sigma}
\newcommand {\del}     {\delta}

\newcommand {\pd}      {\partial}
\newcommand {\pdi}     {\pd_i}

\newcommand {\pj}      {\pd_j}
\newcommand {\pk}      {\pd_k}
\newcommand {\pt}      {\pd_t}
\newcommand {\pr}      {\prime}
\newcommand {\cu}      {{\cal U}}

\newcommand {\ch}      {{\cal H}}

\newcommand {\cx}      {X}
\newcommand {\cp}      {P}
\newcommand {\chns}    {\ch_{NS}}

\newcommand {\che}     {\ch_E}

\newcommand {\intx}    {\int d^3x}

\newcommand {\onehalf} {\frac{1}{2}}
\newcommand {\vect}[1] {\mathbf{#1}}
\newcommand {\bx}      {\vect{x}}
\newcommand {\ba}      {\vect{\cx}}
\newcommand {\bb}      {\vect{\cp}}

\newcommand {\bv}      {\vect{v}}
\newcommand {\bq}      {\vect{q}}

\newcommand {\bn}      {\vect{\nabla}}

\begin{document}

\title{Scales of a fluid}

\author{Billy D. Jones} \email{bdjwww@uw.edu}
\affiliation{Applied Physics Laboratory, University of Washington, Seattle, WA 98105}
\date{2014-09-29}

\begin{abstract} 

The flow of a viscous fluid is perturbed by its internal friction which 
generates heat and leads to a small temperature change. This does not 
occur for an ideal fluid. We would like to resolve this picture as a 
function of the dynamical macroscopic scales of both problems. In order 
to do this we will study the evolution of the Navier-Stokes Hamiltonian with 
the classical similarity renormalization group in the region of small 
viscosity. The connection between the Euler and Navier-Stokes fluids will 
be pursued, but also the viscous structures that arise will be studied 
in their own right to determine the low-order velocity correlators of 
realistic fluids such as single-component air and water. 
The canonical coordinate of the Navier-Stokes Hamiltonian is a 
vector field that stores the initial position of all the fluid particles. 
Thus these appear to be natural coordinates for studying arbitrary 
separations of fluid particles over time. This connection will be pursued 
and the region where the classic 1926 Richardson 4/3 scaling law holds 
will be determined. The evolution of the Euler Hamiltonian will also be studied 
and we will attempt to map its singular structures to those of the 
small-viscosity Navier-Stokes fluid.

\end{abstract}

% -----------------------------------------------------------------------
%47. Fluid dynamics
% -----------------------------------------------------------------------
%47.10.-g	General theory in fluid dynamics
% ------------------------------------------
%47.10.ad Navier-Stokes equations
%47.10.Df Hamiltonian formulations
% ------------------------------------------
%47.11.-j	Computational methods in fluid dynamics
%47.11.St	Multi-scale methods
% -----------------------------------------------------------------------
%05. Statistical physics, thermodynamics, and nonlinear dynamical systems
% -----------------------------------------------------------------------
%05.10.Cc	Renormalization group methods
% -----------------------------------------------------------------------
\maketitle
\thispagestyle{empty}

%\tableofcontents
%\thispagestyle{empty}
%\newpage

\setcounter{page}{1}

\section{Introduction}

%\vspace{-6pt}
The Navier-Stokes and Euler equations of fluid dynamics 
apparently do not map smoothly onto each other in the limit of vanishing viscosity:
the zero viscosity and infinitesimal viscosity fluids do not appear to be limits of the same theory. 
Singular velocity gradients are the culprit, but we would like to understand this connection better.
We therefore focus on the energy dissipation aspect of the Navier-Stokes equation in 
local thermodynamic equilibrium. The dissipation term of the heat equation due to the internal  
friction (viscosity) of a fluid is just another volume heat source that increases the temperature of the fluid slightly,
however it is still a closed thermodynamic system and can be studied with Hamiltonian techniques. 
Thus, the Navier-Stokes Hamiltonian is derived from first principles including the nonholonomic 
entropy constraint and it is shown that the 
dynamical coordinate of a dissipative fluid is a vector field that stores the initial position of all 
the fluid particles---these appear to be natural coordinates 
for studying arbitrary separations of fluid particles over time. 

The Euler and Navier-Stokes Hamiltonians are used to compare 
the vanishing viscosity limit between the two theories. 
It is shown that they have the same number of degrees of freedom in three spatial dimensions: 
six independent scalar potentials; but in the Navier-Stokes 
case, the potentials are actually two vector fields that are canonical coordinate-momentum field pairs. 
Thus, in these ``coordinates'' it is easy to see that the two theories 
have different dynamical degrees of freedom (fields). We will study this connection further
to understand its dynamical consequences better. A final motivation for using
Hamiltonian field theory techniques is that they allow convenient approximations and 
can be used to systematically integrate the equations 
of motion of a fluid one scale at a time with the 
similarity renormalization group which has been shown to be fruitful in nuclear 
and condensed matter physics. 
The connection to classical physics follows from the canonical Poisson bracket structure of 
the fields of the Navier-Stokes Hamiltonian along with its Poisson bracket   
with an arbitrary classical dissipative observable. Thus the stage has been set to 
study the scales of a classical fluid (see Fig.~1) and to better understand the connection between 
the Euler and Navier-Stokes theories in the limit of vanishing but nonzero viscosity---heat 
and eventually diffusion will matter in this work. 

%%%%%%%%%%%%%%%%%%%%%%%%%%% figure picture
\vspace{12pt}
\begin{figure}[!ht]
\centering
    \includegraphics[scale=0.75]{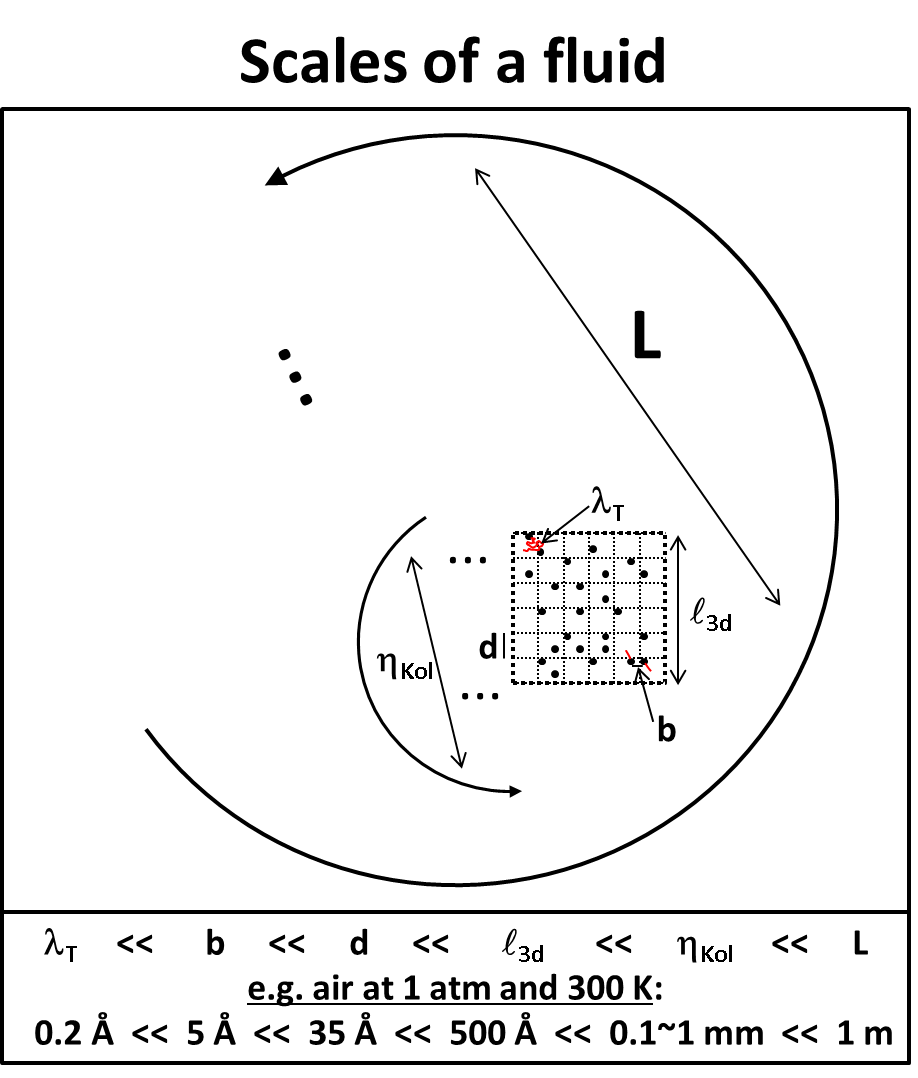} % largest 0.9, smallest 0.5
		%\vspace{-12pt}
\caption{A typical fluid eddy of size $L$ with its macroscopic and microscopic substructure. 
A Kolmogorov microscale eddy of size $\eta_{Kol}$ is the smallest macroscopic structure. 
The successive microscopic substructure is the mean free path $\ell_{3d}$, the average molecular separation $d$, 
the molecular impact parameter $b$ and the thermal wavelength $\lam_T$. For air at room temperature, all of these 
scales are separated by at least an order of magnitude as shown. 
For water, the story is similar but then $b$, $d$, and $\ell_{3d}$ are all of the same order of magnitude  
with the microscopic degrees of freedom strongly coupled. 
Nevertheless in both cases (air and water) $\lam_T$ is much smaller than 
the average separation between the molecules themselves and their quantum nature is therefore inaccessible. 
In addition, in both cases, the mean free path is much smaller than dissipation scale 
$\eta_{Kol}$ and macroscale $L$; therefore the continuum approximation is valid and 
the Navier-Stokes equation becomes the paradigm of interest.}
\label{fig:picture}
\end{figure}
%%%%%%%%%%%%%%%%%%%%%%%%%%%%%%%

\section{Dynamics of a fluid}

The dynamics of a nonrelativistic classical single-component fluid are given by the equations of 
motion for its velocity, density and entropy fields \cite{landau}:
\bee
\rho D_t v_i = -\pdi p + \pj\sig^\pr_{ij}\;,~~
D_t\rho=-\rho\bn\cdot\bv\;,~~
\rho T D_t s=\sig^\pr_{ij}\pj v_i-\bn\cdot\bq\;,
\eee
where $D_t=\pt+\bv\cdot\bn$ is the convective derivative with 
$\pt$ being a shorthand for $\frac{\pd}{\pd t}$. Repeated indices are 
summed over the three spatial dimensions and $\pdi$ is a shorthand for $\frac{\pd}{\pd x_i}$. 
$p$ is pressure, $T$ is temperature, $\sig^\pr_{ij}$ is the viscous stress tensor 
and $\bq$ is the heat flux defined as  
$\bq=-\kappa\,\bn T$ where $\kappa$ is the thermal conductivity. The viscous stress tensor 
for a Newtonian fluid (which defines the Navier-Stokes equation---the first relation above for $D_t v_i$) is given by
\bee
\sig^\pr_{ij}\equiv\eta (\pdi v_j+\pj v_i)+\zeta^\pr\del_{ij}\bn\cdot\bv\;,
\eee
where $\eta$ and $\zeta$ are the shear and bulk viscosity respectively 
and $\zeta^\pr\equiv\zeta-2\eta/3$---defined 
such that the trace of the viscous stress tensor is independent of shear viscosity. 

We carefully wrote these complete equations of motion to be concrete: this is 
what we mean by the ``Navier-Stokes paradigm'' mentioned in Fig.~1. The Navier-Stokes 
Hamiltonian written below reproduces these equations of motion exactly as shown 
in \cite{Hns,Hiroki}. The entropy equation of motion above is known as the heat equation 
and is a nonholonomic (path-dependent) constraint \cite{Goldstein} on the motion of the fluid. Note how 
the viscous stress tensor, $\sig^\pr_{ij}$, is in both the heat equation and the Navier-Stokes equation 
itself: this is how the heat generated by the internal friction of the 
fluid gets coupled into its motion and causes the 
temperature to rise in its own wake so to say. It is 
a self-energy correction for the system, but energy is still conserved since ``heat'' is included 
in what we mean by energy (the main lesson of thermodynamics).

These equations of motion follow from first principles of momentum, mass 
and energy conservation \cite{landau}.   
As shown in \cite{Hns,Hiroki}, they also follow from the Euler-Lagrange equations of their respective Lagrangian, 
and the Hamilton equations of their respective Hamiltonian. 
We use the Lagrangian to derive the initial Hamiltonian (which gets changed due to 
renormalization) but in what follows to keep the discussion clearer, hereafter we 
only discuss Hamiltonians. 
The interesting thing is that starting from first principles one is led 
to a drastically different form \cite{Hns} for the Hamiltonian of the Euler 
and Navier-Stokes fluids (defined as fluids satisfying the respective equation). 
Note that the equations of motion for the Euler fluid are the same as the above
with $\sig^\pr_{ij}$ set to zero (which includes removing gradients! and is not just setting viscosity to zero).

Here we write the kinematic results for the difference between the Navier-Stokes and Euler Hamiltonian 
degrees of freedom. For this work, we would like to pursue the dynamical consequences of this difference 
using the similarity renormalization group (described in the next section) to evolve the Navier-Stokes Hamiltonian 
from its macroscopic small to large scales ($\eta_{Kol}$ through $L$---see Fig.~1) to obtain the 
velocity correlation functions and 
particle separation distributions of the theory at scale `s' and time `t'.

For the ideal fluid, the Euler Hamiltonian density is given by \cite{Hns,Hiroki,Zakharov}
\bee
\che(\rho, \phi; \beta, \alpha; s, \lam)=\onehalf\rho
\left(\bn\phi-\frac{\alpha}{\rho}\bn\beta-\frac{\lam}{\rho}\bn s\right)^2
+\cu(\rho,s)\;.
\eee
Note that the arguments of $\che$ on the left are 
written in terms of its three dynamical coordinate fields: $\rho$, $\beta$, and $s$; and 
their respective conjugate momentum fields: $ \phi$, $\alpha$, and $\lam$. 
$\rho$ and $s$ are the same density and entropy field as in the equations of motion above.
$\phi$ is the same velocity potential as in potential flow fluid mechanics ($\bv=\bn\phi$). 
$\alpha$ and $\beta$ are a Clebsch-potential pair and $\lam$ is the lagrange multiplier field for the entropy constraint 
of an ideal fluid (i.e. $D_t s = 0$). 
$\cu$ is the internal energy density. Again, see \cite{Hns} for further details, but note that 
a Hamiltonian and its density are related by $H=\intx\,\ch$.
 
For the viscous fluid we are led to a very different Hamiltonian due 
to the dissipation term of the heat equation: $\sig^\pr_{ij}\pj v_i \equiv \rho\,\varepsilon$, 
where $\varepsilon$ is the well-known energy dissipation field which in a region 
of fully-developed turbulent scaling is constant or nearly constant. This dissipation makes 
the entropy constraint nonholonomic and necessitates the introduction of 
vector field $\ba(\bx,t)$ \cite{Hiroki} in a fashion at least reminiscent of gauge invariance \cite{Zakharov}. 
This new field $\ba(\bx,t)$ is just a coordinate transformation, and as shown in \cite{Hiroki, Hns} it is a 
canonical transformation with the $(\bx,\vect{p})$ and $(\ba,\bb)$ pairs having the same Poisson bracket structure \cite{Goldstein}. With all the algebra worked out in \cite{Hns}, the Hamilton equations 
of the following Hamiltonian are equivalent to the equations of motion that led off this section.   
The Navier-Stokes Hamiltonian density is given by \cite{Hiroki}
\bee
\chns\left(\ba,\bb,s(\ba)\right)=\frac{(P_i\bn X_i)^2}{2\,J(\bn\ba)\,\rho_0(\ba)}
+\cu(J(\bn\ba)\,\rho_0(\ba),s(\ba))\;.
\eee
The degrees of freedom of this Hamiltonian are coordinate vector field $\ba(\bx,t)$ and 
its conjugate momentum vector field $\bb(\bx,t)$. 
$s(\ba)$ is the same entropy field as in the heat equation, but here it is given by the following 
nonholonomic field variation constraint \cite{Hiroki,Hns}:
\bee
\frac{\del s}{\del X_i}=\left(\pj s+\frac{\pk\sig_{jk}^\pr}{\rho\,T}\right)\frac{\pd x_j}{\pd X_i}
\;.
\eee
Note that $\chns$ starts out quartic in the fields in the numerator of the first term and 
{\it there is no standard quadratic term of field theory} \cite{Lvov}. $J(\bn\ba)$ is the Jacobian explicitly 
given in \cite{Hns} with six terms in total and each term being 
cubic in $\pj X_i$, and for this first ``$p^2/(2m)$'' term it is in the denominator which
makes its contributions nonlocal.  Finally, $\rho_0(\ba)$ is the initial density 
set by the physics of the problem which  
often allows $\rho = J\rho_0$ to be approximated as a constant or near-constant mean 
and then to perturb about this mean. We would like to explore these ideas further with this work. 

In summary, the Euler and Navier-Stokes fluids have quite different pathlines for their fluid particles given 
by their respective velocity field:  
\beaa
\bv_E&=&\fbox{$\bn\phi-\frac{\alpha}{\rho}\bn\beta-\frac{\lam}{\rho}\bn s$}\;,\\
\bv_{NS}&=&\fbox{$-\frac{P_i}{\rho}\bn X_i$}=-\frac{P_i\bn X_i}{J(\bn\ba)\,\rho_0(\ba)}\;.
\eeaa
Vector field $\ba$ is just another coordinate like $\bx$, so the fact that there are three scalar potential pairs 
in the Navier-Stokes fluid $(X_1,P_1;X_2,P_2;X_3,P_3)$ is directly related to the choice of working in three spatial dimensions. 
The Navier-Stokes fluid seems to be perfectly coupled to three spatial dimensions.
There are two further points to be made both highlighting the differences between the Euler and Navier-Stokes fluids 
even though upon first sight, these $\bv_E$ and $\bv_{NS}$ decompositions look similar. 
First, for $\bv_E$ note how the $\bn\phi$ term has a plus sign and does not have any density dependence whereas
the other two terms have opposite sign and have density dependence. For the Navier-Stokes fluid,   
the signs of all three terms can be made to be the same in $\bv_{NS}$ 
with the conjugate momenta of $\chns$ related by a positive sign: $\pi_{X_i}=+P_i$, and there is no asymmetry 
in the density dependence.   
$\phi$ is of course the standard velocity potential of fluid mechanics and $\bv_E$ is the 
so-called Clebsch decomposition of the velocity field 
with Gauss potential \cite{jackiw} pairs ($\alpha$,$\beta$) and ($\lam$,$s$). Interestingly, to obtain the 
Navier-Stokes Hamiltonian one had to introduce coordinate transformation $\ba(\bx,t)$ 
in order to handle the entropy constraint properly and this already gave enough degrees of freedom and 
so a velocity potential was not required. The Euler and Navier-Stokes fluids are very different. 
Second, $\bv_E$ and $\bv_{NS}$ being different is even more readily apparent 
if we recall that $\rho$ is a dynamical field for the Euler fluid (satisfying the standard continuity equation), 
whereas for the Navier-Stokes fluid, $\rho=J\rho_0$ is a constraint which in terms of vector field $\ba$ 
is quite complex: see \cite{Hns} for an explicit expression for $J(\bn\ba)$. Also note 
that since $J$ (with derivatives of fields) is in the denominator, it is a nonlocal operator in $\chns$. 
These differences should not come as a huge surprise since 
the Euler and Navier-Stokes fluids apparently do not map smoothly onto each other:  
the zero viscosity and infinitesimal viscosity fluids do not appear to be limits of the same theory. 
Perhaps these variational principle forms of the theories help to make this clearer, and 
with this work we would like to pursue the dynamical consequences of this further.

The Poisson bracket structure of the fields of $\chns$ along with its Poisson 
bracket with an arbitrary dissipative observable is derived in \cite{Hns}. This sets 
the stage for the similarity renormalization group (SRG) introduced in the next section.
The connection between nuclear physics (the original arena of the SRG \cite{SRG}) 
and fluid dynamics is made through 
the Poisson bracket structure of the classical Hamiltonian of interest, 
in this case $\chns$. 

\section{Classical similarity renormalization group}

In order to better understand the scales of a classical fluid and the 
connection between the Euler and Navier-Stokes fluids 
we propose studying the physics of a viscous fluid as the viscosity vanishes. As is well known, on dimensional 
grounds $\eta_{Kol}\sim\nu^{3/4}/\varepsilon^{1/4}$, therefore this entails understanding 
the physics of a viscous fluid at it smallest macroscopic scales. We propose using the 
similarity renormalization group (SRG) to study the flow of the Navier-Stokes Hamiltonian 
from the smallest macroscopic scales, around $\eta_{Kol}$, (see Fig.~1) where the energy is dissipated 
through the largest scales, around $L$, where the energy is input. 

The classical SRG acting on a Hamiltonian at scale $s$ 
(which can be thought of as the spatial ``$s$ for size'' of the region under study) 
is given by the following flow equation
\bee
\frac{dH_s}{ds}=-\left[\eta_s,H_s\right]_{P.B.}\;,~~
H_s\equiv H_0+V_s\;,~~
\eta_s=[H_0,V_s]_{P.B.}\;,
\eee
where `P.B' implies `Poisson bracket' and $\eta_s$ is the generator of this scale transformation.
This is a new result based on the correspondence principle \cite{Sakurai} 
applied to Wegner's flow equation \cite{Wegner}. Since it is a new result,
we would like to show that the sign of this equation (which came from $i^2=-1$) is correct 
by comparing a simple fixed-source calculation in quantum field theory \cite{fixedsource} with an analogous 
classical Hamiltonian: a harmonic oscillator with a linear potential due to, for example, gravity.
Thus, say the starting classical Hamiltonian with coordinate $q$ and momentum $p$ is given by
\bee
H = \frac{p^2}{2}+\frac{q^2}{2}+ g\,q\;,
\eee
with coupling $g$ (here the acceleration of gravity with unit mass and spring constant).
In order to integrate out the effects of gravity (e.g.\ with the viscous fluid problem we could choose 
the dissipation operator of the entropy constraint here), next we choose
\bee
V = g\,q\;,
\eee
which implies that the free Hamiltonian is $H_0 = \frac{p^2}{2}+\frac{q^2}{2}$. Sticking this $H_0$ 
and $V$ into the above classical SRG flow equation gives 
(hereafter we drop the designator ``P.B'', but it is implied in all this work)
\beaa
[H_0,V]&=&\left[\frac{p^2}{2}+\frac{q^2}{2},g\,q\right]=-g\,p\;,\\
\frac{dH}{ds}&=&-\left[-g\,p\,,\,\frac{p^2}{2}+\frac{q^2}{2}+ g\,q\right]=-g\,q-g^2\;,
\eeaa
using the standard rules of a Poisson bracket \cite{Goldstein}. This last line is the result that corresponds exactly 
with like terms in the fixed-source quantum field theory problem \cite{fixedsource}. Like there, in order to have 
a Hamiltonian with fixed structure as it runs with scale $s$, we change the initial Hamiltonian to the following 
ansatz, with a new ``self-energy'' term and a running coupling $g_s$:
\bee
H_s = \Sigma_s+\frac{p^2}{2}+\frac{q^2}{2}+ g_s\,q\;.
\eee
Requiring consistency with this above $dH/ds$ result gives the following running coupling equations
\bee
\frac{dg_s}{ds}=-g_s\;,~~
\frac{d\Sigma_s}{ds}=-g_s^2\;,
\eee
with solution
\bee
g_s=g_0\,e^{-s}\;,~~
\Sigma_s=\Sigma_{s0}-\frac{g_0^2}{2}\left[e^{-2s_0}-e^{-2s}\right]\;.
\eee
The initial scale is $s=0$ and as $s\rightarrow\infty$ we see the gravity interaction is integrated out, 
with the self-energy $\Sigma_s$ dressed with a background field like in the Yukawa meson-cloud problem \cite{fixedsource}, 
but here derived completely in the context of classical physics. The parallelism between the analogy is quite striking 
showing the correct sign identification for the above classical SRG flow equation.

\section{Summary}

%\vspace{-6pt}
The prime directive of this work is to understand the renormalization of the 
Navier-Stokes Hamiltonian through the classical similarity renormalization group to help
elucidate the connection between the ideal and viscous theories. We will 
seek the connection first from studying the small-scale macroscopic structures of the viscous theory near the 
Kolmogorov scale as the dissipation operator is run with similarity scale $s$. 
Since the Navier-Stokes Hamiltonian dynamical coordinate is a vector field that stores the initial 
positions of all the fluid particles, we propose the study of its low-order velocity correlators 
and particle separation distributions as a function of scale, 
in order to determine the region for which the classic Richardson 4/3 scaling law holds \cite{turbvoyage}. 
If the correct scales are resolved, the mechanism for the internal friction 
giving rise to heat and motion should be more readily apparent. 
The classical similarity renormalization group 
flow equation and Navier-Stokes Hamiltonian could shed new light on an old problem.

\end{document}